\documentclass[preprint,authoryear,12pt]{elsarticle}

\usepackage{floatrow}
\usepackage{amsmath}
\usepackage{amssymb}

\journal{arXiv}

\usepackage{amsmath}
\usepackage{amsthm}

\newtheorem*{theorem*}{Theorem}
\newtheorem{lemma}{Lemma}
\newtheorem*{lemma*}{Lemma}
\newtheorem{remark}{Remark}

\usepackage{optidef}
\usepackage{subcaption}

\allowdisplaybreaks

\usepackage{todonotes}

\usepackage{calc}
\usepackage{algorithm}
\usepackage{algpseudocode}
\usepackage{enumitem}
\usepackage{url}

\usepackage{hyperref}
\usepackage{breakurl}



\usepackage{mathtools}

\mathtoolsset{showonlyrefs}

\begin{document}

\begin{frontmatter}



\title{Combining independent $p$-values in replicability analysis: A comparative study}

\author{Anh-Tuan Hoang}
\author{Thorsten Dickhaus\corref{cor1}}
\cortext[cor1]{Institute for Statistics, University of Bremen, 
P. O. Box 330 440, 28344 Bremen, Germany. Tel: +49 421 218-63651. E-mail address: dickhaus@uni-bremen.de (Thorsten Dickhaus).}

\address{Institute for Statistics, University of Bremen, Germany}

\begin{abstract}
Given a family of null hypotheses $H_{1},\ldots,H_{s}$, we are interested in the hypothesis $H_{s}^{\gamma}$ that at most $\gamma-1$ of these null hypotheses are false. Assuming that the corresponding $p$-values are independent, we are investigating combined $p$-values that are valid for testing $H_{s}^{\gamma}$. In various settings in which $H_{s}^{\gamma}$ is false, we determine which combined $p$-value works well in which setting. Via simulations, we find that the Stouffer method works well if the null $p$-values are uniformly distributed and the signal strength is low, and the Fisher method works better if the null $p$-values are conservative, i.e. stochastically larger than the uniform distribution. The minimum method works well if the evidence for the rejection of $H_{s}^{\gamma}$ is focused on only a few non-null $p$-values, especially if the null $p$-values are conservative. Methods that incorporate the combination of $e$-values work well if the null hypotheses  $H_{1},\ldots,H_{s}$ are simple. 
\end{abstract}

\end{frontmatter}



\section{Introduction}
\label{sec:introduction}

Given a set of studies, which are examining related research hypotheses under different conditions, it is often of interest to assess whether findings can be made in at least $\gamma\geq 2$ of the considered studies. Studies may, for example, differ in their population or laboratory methods. The search for results in at least two studies is called replicability analysis. It alleviates the possibility that a positive outcome depends on the specific settings of a single study. 

Formally, we consider a family of $s\geq 2$ null hypotheses $H_{1},\ldots,H_{s}$ and their corresponding alternative hypotheses $K_{1},\ldots,K_{s}$. For each pair of hypotheses $H_{i}$ and $K_{i}$ we assume that a $p$-value $p_{i}$ is available and that these $p$-values $p_{1},\ldots,p_{s}$ are jointly stochastically independent. For $\gamma\leq s$, we are interested in the partial conjunction/replicability null hypothesis
\begin{equation}
\label{eq:H}
H_{s}^{\gamma}=\{\text{at least $s-\gamma+1$ of the null hypotheses $H_{1},\ldots,H_{s}$ are true}\},
\end{equation} 
thus its alternative is that at least $\gamma$ null hypotheses are false. Our goal is to compare different $p$-value combinations for $p_{1},\ldots,p_{s}$ that are valid for $H_{s}^{\gamma}$. A $p$-value is called valid for a null hypothesis $H$ if it is stochastically not smaller than the uniform distribution on $[0,1]$ ($\mathrm{Uni}[0,1]$) under all parameter values that entail validity of $H$.

Since it holds $H_{s}^{1}\subseteq H_{s}^{2}\subseteq\cdots\subseteq H_{s}^{s}$, valid $p$-values for $H_{s}^{1}$ need not be valid for $H_{s}^{\gamma}$, $\gamma\geq 2$. \cite{benjamini2008screening} investigated the theory of testing the partial conjunction null hypotheses $H_{s}^{\gamma}$. They show that valid $p$-values for $H_{s}^{\gamma}$ can be derived from combination $p$-values for $H_{s-\gamma+1}^{1}$ by essentially combining the $s-\gamma+1$ largest $p$-values.

There are several ways to combine the independent $p$-values $p_{1},\ldots,p_{s}$ of a set of null hypotheses $H_{1},\ldots,H_{s}$ to test for the null hypothesis $H_{s}^{1}$. \cite{birnbaum1954combining} showed that for each $p$-value combination that is non-decreasing in each $p$-value, there exists an alternative hypothesis for which the combination is best. Nevertheless, we consider different null hypothesis setups, and identify which of our considered combinations work best in which general situation. 

A common way of combining $p$-values for $H_{s}^{1}$ is via averaging, see also \cite{vovk2020combining}. This approach evaluates the sum of the transformed $p$-values $f(p_{1},\ldots,p_{s})=\sum_{i}\varphi(p_{i})$. If the distribution of $f(U_{1},\ldots,U_{s})$ is known with cdf $F_{f}$, where $U_{i}\sim\mathrm{Uni}[0,1]$, $i=1,\ldots,s$, then defining $F_{f}(f(p_{1},\ldots,p_{s}))$ leads to a valid $p$-value under $H_{s}^{1}$. Two well known examples of such are Fisher's method which uses $\varphi=\mathrm{log}$, and Stouffer's method, which uses $\varphi(x)=-\Phi^{-1}(1-x)$, where $\Phi^{-1}$ is the quantile function of the standard normal distribution on $\mathbb{R}$. If the distribution of $f(U_{1},\ldots,U_{s})$ is unknown, one can instead modify $p=G_{0}(f(p_{1},\ldots,p_{s}))$ with a suitable function $G_{0}$ so that $p$ is at least valid under $H_{s}^{1}$. This includes for example the arithmetic mean and the harmonic mean, cf. \cite{ruschendorf1982random}, \cite{vovk2020combining}, \cite{wilson2019harmonic}. Either way, large $p$-values can overshadow small $p$-values in averaging methods, which can be problematic if the null $p$-values are conservative under nulls, that is, if they are stochastically larger than $\mathrm{Uni}[0,1]$. On the other hand, none of the $p$-values need to be smaller than a significance level $\alpha\in(0,1)$ for the combined $p$-value to be smaller than $\alpha$. Thus, averaging methods can be powerful if the evidence for a rejection of the global null hypothesis is spread out between the $p$-values. Pearson's method of averaging via a product of the $p$-values is of a similar nature, cf. \cite{pearson1938probability}.

On the other hand, there are $p$-value combination functions that do not take the size of all $p$-values fully into account. For example the combined $p$-value $s\cdot\mathrm{min}\{p_{1},\ldots,p_{s}\}$ resulting from the Bonferroni method is relatively unaffected by conservative $p$-values, but the $p$-value cannot be smaller than $\alpha$ if none of the marginal $p$-values are. Similarly, since the minimum $\mathrm{min}\{p_{1},\ldots,p_{s}\}$ of stochastically independent $\mathrm{Uni}[0,1]$-distributed $p$-values is $\mathrm{Beta}$-distributed with parameters $1$ and $s$, $\mathrm{Beta}(1,s)$, we can also consider $p=F_{\mathrm{Beta}(1,s)}(\mathrm{min}\{p_{1},\ldots,p_{s}\})$ as a valid $p$-value for $H_{s}^{1}$. Another example is the maximum of the $p$-values, which is valid for $H_{s}^{1}$ if the $p$-values are independent, cf. \cite{vovk2020combining}. 

Similarly to the work of \cite{loughin2004systematic}, we differentiate between alternative hypotheses that have minimally spread evidence and ones that have spread out evidence among all the false null hypotheses. If $s=6$ and $\gamma=2$, the null hypothesis $H_{s}^{\gamma}$ is, for example, false if only two null hypotheses are false or if all null hypotheses are false. However, \cite{loughin2004systematic} only considered the global null hypotheses $H_{s}^{1}$, that every null hypothesis is true. We extend his work by taking the more general partial conjunction hypothesis $H_{s}^{\gamma}$ into account. \cite{birnbaum1954combining} already noted that in case of $s=2$ studies, the Wilkinson $p$-value (case $1$) is more sensitive to evidence in one study than the Fisher $p$-value, cf. \cite{wilkinson1951statistical}. Furthermore, \cite{loughin2004systematic} only considered $\mathrm{Uni}[0,1]$-distributed null $p$-values. However, it is known that for example in case of composite null hypotheses, conservative null $p$-values are more common, cf. \cite{hoang2020usage}, \cite{hoang2021randomized}. In simulations, we investigate how well the different $p$-value combination functions deal with conservative null $p$-values.

Not covered in this paper is the kind of meta analysis that tests $H_{s}^{1}$ against the alternative of $H_{s}^{s}$, which is a proper subset of the alternative of $H_{s}^{1}$ if $s>1$, i.e. each of the null hypotheses $H_{1},\ldots,H_{s}$ are either all true or all false (for example repetition of an experiment). \cite{kocak2017meta} modeled the marginal $p$-values under alternatives as $\mathrm{Beta}(\alpha,\beta)$-distributed, Beta-distributed with parameters $\alpha$ and $\beta$, and determined which $p$-value combinations work well for this question in which subsets of $(0,\infty)^{2}$ for the parameters $(\alpha,\beta)$. Under the assumption that the marginal $p$-values are $\mathrm{Uni}[0,1]$-distributed under $H_{s}^{1}$, \cite{heard2018choosing} calculate $p$-value combinations as likelihood ratio tests and therefore uniformly most powerful test statistics for the above kind of meta analysis under several models.

This work is structured as follows. In Section$~\ref{sec:modelsetup}$ we introduce our model and notations, and in Section$~\ref{sec:pvaluecombinations}$ we present the $p$-value combinations that we consider. Section$~\ref{sec:simulations}$ contains our comparisons of the $p$-value combination functions via simulations. Finally, we conclude with a discussion in Section$~\ref{sec:discussion}$.


\section{Model Setup}
\label{sec:modelsetup}

Let $\left(\Omega,\mathcal{F},(\mathbb{P}_{\boldsymbol{\theta}})_{\boldsymbol{\theta}\in\Theta}\right)$ be a statistical model, and let $\boldsymbol{X}$ be the data and $\boldsymbol{\theta}\in\Theta$ the parameter of the model and $\Theta$ the corresponding parameter space. We consider a set of null hypotheses $\left(H_{i}\right)_{i=1,\ldots,s}$ and their corresponding alternatives $\left(K_{i}\right)_{i=1,\ldots,s}$ such that $H_{i}$ and $K_{i}$ are non-empty subsets of the parameter space $\Theta\subseteq\mathrm{\mathbb{R}}^{s}$. We assume that $H_{i}=\left\{\boldsymbol{\theta}:\theta_{i}\leq 0\right\}$ and $K_{i}=\left\{\boldsymbol{\theta}:\theta_{i}>0\right\} $ for each $i=1,\ldots,s$. Thus, each hypothesis pair $H_{i}$ and $K_{i}$ only depends on the $i$-th component of the parameter value $\boldsymbol{\theta}$, $i=1,\ldots,s$.

Let a set of corresponding $p$-values $\left(p_{i}\right)_{i=1,\ldots,s}$ be given, such that for any parameter value $\boldsymbol{\theta}=(\theta_{1},\ldots,\theta_{s})^{T}\in\Theta$
\begin{equation*}
p_{i}(\boldsymbol{X})\sim f_{\theta_{i}},\;i=1,\ldots,s,
\end{equation*}
where $f_{\theta_{i}}$ is a Lebesgue density with support on $\left[0,1\right]$. More particularly, we assume that the density function $f_{\theta_{i}}$ only depends on the $i$-th component $\theta_{i}$ of $\boldsymbol{\theta}$. Throughout this paper, we use $p_{i}\equiv p_{i}(\boldsymbol{X})$ by abuse of notation, and say '$p$-value' and '$p$-variable' interchangeably.

We make the following general assumptions to our model:

\begin{description}
\item[$(A1)$] The $p$-values $p_{1},\ldots,p_{s}$ are jointly stochastically independent under each parameter value $\boldsymbol{\theta}\in\Theta$.
\item[$(A2)$] Under any $\boldsymbol{\theta}\in\Theta$ such that $\theta_{i}=0$, we assume that $f_{0}\left(t\right)=\boldsymbol{1}\left\{0\leq t\leq 1\right\}$, i.e. that $p_{i}(\boldsymbol{X})$ is $\mathrm{Uni}[0,1]$-distributed under $\theta_{i}=0$, $i=1,\ldots,s$. 
\item[$(A3)$] For each $i$, we assume that the $p$-value $p_{i}(\boldsymbol{X})$ is stochastically decreasing in $\theta_{i}$, i.e. $p_{i}(\boldsymbol{X})^{(\theta_{i})}\leq_{\mathrm{st}}p_{i}(\boldsymbol{X})^{(\tilde{\theta}_{i})}$ if and only if $\theta_{i}\geq\tilde{\theta}_{i}$. 
\end{description}

Assumption $(A1)$ is for example fulfilled if the null hypotheses $H_{1},\ldots,H_{s}$ are from a set of independent studies. If the parameter space $\Theta$ contains no parameter values $\boldsymbol{\theta}$ with negative $i$-th components $\theta_{i}$, $p_{i}$ is $\mathrm{Uni}[0,1]$-distributed under each $\boldsymbol{\theta}\in H_{i}$. Otherwise, $p_{i}$ may be conservative if $\theta_{i}<0$, $i=1,\ldots,s$.

The relation $\leq_{\mathrm{st}}$ in assumption $(A3)$ denotes the usual stochastic order between two random variables, cf. for example \cite[Chapter$~1.A$]{shaked2007stochastic}. The notation $p_{i}(\boldsymbol{X})^{(\theta_{i})}$ refers to the distribution of $p_{i}(\boldsymbol{X})$ under $\theta_{i}$. Under assumptions $(A2)$ and $(A3)$, $p_{i}$ is a valid $p$-value for $H_{i}$ and parameters $\boldsymbol{\theta}$ with $\theta_{i}=0$ are the least favorable parameter configurations (LFC parameters). See for example Section$~2$ in \cite{hoang2021randomized} for a definition of LFC-based $p$-values.

\begin{remark}
\label{rm:likelihood}
Assumption $(A3)$ is for example fulfilled if $p_{i}$ is an antitone transformation of a test statistic $T_{i}(\boldsymbol{X})$ such that $(T_{i}(\boldsymbol{X})^{(\theta_{i})})_{\theta_{i}}$ is likelihood ratio ordered, that is, if the distribution of $T_{i}(\boldsymbol{X})$ under $\boldsymbol{\theta}$ is smaller under the likelihood ratio order than under $\tilde{\boldsymbol{\theta}}$ if and only if $\theta_{i}\leq\tilde{\theta}_{i}$ holds for their $i$-th components (cf. for example Chapter$~1.C$ in \cite{shaked2007stochastic} for a definition of the likelihood ratio order).
\end{remark}

We are interested in the partial conjunction null hypothesis $H_{s}^{\gamma}$ from \eqref{eq:H}, where $1\leq\gamma\leq s$ is a given constant. The goal of this work is to compare $p$-value combination maps $f:\left[0,1\right]^{s}\rightarrow\left[0,1\right]$ for which $f(p_{1},\ldots,p_{s})$ is a valid $p$-value for $H_{s}^{\gamma}$.

\section{Combination functions for \texorpdfstring{$p$}{p}-values}
\label{sec:pvaluecombinations}

In this section, we introduce the $p$-value combinations $f(p_{1},\ldots,p_{s})$ that we investigate for the null hypothesis $H_{s}^{\gamma}$. Let $U_{1},\ldots,U_{s}$ be stochastically independent and identically $\mathrm{Uni}[0,1]$-distributed random variables. 

We first assume the existence of a $p$-value combination function $g:\left[0,1\right]^{s-\gamma+1}\rightarrow\left[0,1\right]$, that is non-decreasing in each argument and valid for the null hypothesis $H_{s-\gamma+1}^{1}$, i.e. 
\begin{equation}
\label{eq:f0validity}
\mathrm{Uni}[0,1]\leq_{\mathrm{st}}g\left(U_{1},\ldots,U_{s-\gamma+1}\right).
\end{equation}
Let $p_{(1)}\leq\ldots\leq p_{(s)}$ be the ordered $p$-values. According to Lemma $1$ in \cite{benjamini2008screening}, this combination function applied to the $s-\gamma+1$ largest $p$-values $p_{(\gamma)},\ldots,p_{(s)}$ among $p_{1},\ldots,p_{s}$ is valid for $H_{s}^{\gamma}$, i.e. $f\left(p_{1},\ldots,p_{s}\right)$ is valid for $H_{s}^{\gamma}$, where $f:\left[0,1\right]^{s}\rightarrow\left[0,1\right]$ is a combination function with
\begin{equation}
f\left(p_{1},\ldots,p_{s}\right)=g\left(p_{(\gamma)},\ldots,p_{(s)}\right).
\end{equation}

Hence, in order to find $p$-value combination functions for the partial conjunction null hypothesis $H_{s}^{\gamma}$, we only have to consider $p$-value combination functions $g:\left[0,1\right]^{s-\gamma+1}\rightarrow\left[0,1\right]$ for $H_{s-\gamma+1}^{1}$. The functions $g$, that we use in this paper, can be divided into two classes.
\begin{description}
\item[$1.$] We have a component-wise non-increasing function $g_{0}:\left[0,1\right]^{s-\gamma+1}\rightarrow\mathbb{R}$, such that the distribution of $g_{0}\left(U_{1},\ldots,U_{s-\gamma+1}\right)$ is known with continuous cdf $F_{g_{0}}$. We then define the $p$-value $g\left(p_{1},\ldots,p_{s-\gamma+1}\right)=1-F_{g_{0}}\left(g_{0}\left(p_{1},\ldots,p_{s-\gamma+1}\right)\right)$. Note, that it holds $g\left(U_{1},\ldots,U_{s-\gamma+1}\right)\sim\mathrm{Uni}\left[0,1\right]$ by the principle of probability integral transform.
\item[$2.$] For a component-wise non-decreasing function $g_{0}:\left[0,1\right]^{s-\gamma+1}\rightarrow\mathbb{R}$ we consider $g_{0}\left(p_{1},\ldots,p_{s-\gamma+1}\right)$ and find a constant $c\in\mathbb{R}$ such that $\mathrm{Uni}\left[0,1\right]\leq_{\mathrm{st}}c\cdot g_{0}\left(U_{1},\ldots,U_{s-\gamma+1}\right)$. We then define the $p$-value $g\left(p_{1},\ldots,p_{s-\gamma+1}\right)=c\cdot g_{0}\left(p_{1},\ldots,p_{s-\gamma+1}\right)$.
\end{description}

In the following, we take some well known $p$-value combination functions $g$ for $H_{s-\gamma+1}^{1}$ from previous literature. Firstly, as already mentioned in the Introduction, we consider the Fisher and the Stouffer combination. The combined $p$-value by Fisher for $H_{s-\gamma+1}^{1}$ applied to $p_{(\gamma)},\ldots,p_{(s)}$ is defined as
\begin{equation*}
g(p_{(\gamma)},\ldots,p_{(s)})=1-F_{\chi_{2(s-\gamma+1)}^{2}}\left(-2\sum_{i=\gamma}^{s}\mathrm{log}\left(p_{(i)}\right)\right),
\end{equation*} 
where $F_{\chi_{2(s-\gamma+1)}^{2}}$ is the cdf of the $\chi^{2}$-distribution
with $2(s-\gamma+1)$ degrees of freedom (cf. \cite[Section$~21.1$]{fisher1934statistical}). This combination function uses the fact that $f_{0}(U_{1},\ldots,U_{s-\gamma+1})=-2\sum_{i=1}^{s-\gamma+1}\mathrm{log}\left(U_{i}\right)$ is chi-square distributed with $2(s-\gamma+1)$ degrees of freedom.

The combined $p$-value by Stouffer for $H_{s-\gamma+1}^{1}$ applied to $p_{(\gamma)},\ldots,p_{(s)}$ is defined as
\begin{equation*}
g(p_{(\gamma)},\ldots,p_{(s)})=1-\Phi\left(\frac{1}{\sqrt{s-\gamma+1}}\sum_{i=\gamma}^{s}\Phi^{-1}\left(1-p_{(i)}\right)\right). 
\end{equation*} 
where $\Phi$ is the cdf of the standard normal distribution on $\mathbb{R}$ (cf.
\cite[Footnote 14 in Section V of Chapter 4]{stouffer1949american}). This combination function uses the fact that $g(U_{1},\ldots,U_{s-\gamma+1})=(s-\gamma+1)^{-1/2}\sum_{i=1}^{s-\gamma+1}\Phi^{-1}\left(1-U_{i}\right)$ is standard normally distributed. Both combined $p$-values require the $p$-values $p_{1},\ldots,p_{s}$ to be stochastically independent.


The next two combined $p$-values evaluate only the smallest $p$-value. The combined $p$-value using the minimum is defined by 
\begin{equation*}
g(p_{(\gamma)},\ldots,p_{(s)})=F_{\mathrm{Beta}(1,s-\gamma+1)}\left( p_{(\gamma)}\right),
\end{equation*}
where $F_{\mathrm{Beta}(1,s-\gamma+1)}$ is the cdf of the $\mathrm{Beta}$-distribution with parameters $1$ and $s-\gamma+1$. It requires that the $p$-values are stochastically independent, and is motivated by the fact that $g(U_{1},\ldots,U_{s-\gamma+1})=\mathrm{min}\left\{U_{1},\ldots,U_{s-\gamma+1}\right\}$ is $\mathrm{Beta}(1,s-\gamma+1)$-distributed. The Bonferroni method, which utilizes the Bonferroni inequality, leads to 
\begin{equation*}
g(p_{(\gamma)},\ldots,p_{(s)})=(s-\gamma+1)p_{(\gamma)}. 
\end{equation*}
It also evaluates the minimum but does not require independent $p$-values.

Some further $p$-value combination functions that we consider make use of so-called $e$-values (see \cite{grunwald2020safe,vovk2019values}). Their relation to $p$-values is roughly inverse, where higher $e$-values entail stronger evidence against the null. In our simulations in Section$~\ref{sec:simulations}$, we calculate a Bayes factor $e_{j}$ for each null hypothesis $H_{j}$, $j=1,\ldots,s$. These are in some cases $e$-values, i.e. random variables with expected values not greater than one under $H_{j}$. More details on this problem are provided in Section$~\ref{subsec:nullsimulations}$ and in Appendix$~A.1$ of \cite{vovk2020combining}.   

Analogously to the problem of $p$-values, we define a combination function $h$ for $H_{s}^{\gamma}$ by 
\begin{equation*}
h(e_{1},\ldots,e_{s})=h_{0}(e_{(1)},\ldots,e_{(s-\gamma+1)}),
\end{equation*}
where $h_{0}$ is a valid combination function for $H_{s-\gamma+1}^{1}$, i.e. $h_{0}(e_{1},\ldots,e_{s-\gamma+1})$ is a valid $e$-value for $H_{s-\gamma+1}^{1}=\bigcap_{i=1}^{s-\gamma+1}H_{i}$ if $e_{1},\ldots,e_{s-\gamma+1}$ are valid $e$-values for $H_{1},\ldots,H_{s-\gamma+1}$, respectively. We explain in the appendix why $h$ is a valid combination function for $H_{s}^{\gamma}$. Finally, to compare the $e$-value approaches to the ones utilizing $p$-values, we transform the $e$-value $h(e_{1},\ldots,e_{s})$ to a $p$-value, $\mathrm{max}\{h(e_{1},\ldots,e_{s})^{-1},1\}$, for $H_{s}^{\gamma}$ (where $0^{-1}=1$ and $\infty^{-1}=0$).

Some examples of $e$-value combination functions $h_{0}$ for $H_{s-\gamma+1}^{1}$ include the arithmetic mean given by 
\begin{equation*}
h_{0}\left(e_{1},\ldots,e_{s-\gamma+1}\right)=\frac{1}{s-\gamma+1}\sum_{i=1}^{s-\gamma+1}e_{i},
\end{equation*}
and the product given by
\begin{equation*}
h_{0}\left(e_{1},\ldots,e_{s-\gamma+1}\right)=\prod_{i=1}^{s-\gamma+1}e_{i},
\end{equation*}
(cf. \cite{vovk2019values}). Some reasoning on why we chose these functions for $h_{0}$ is given in Propositions$~3.1$ and $4.2$ in \cite{vovk2020combining}.

\section{Simulations}
\label{sec:simulations}

In this section we compare the the $p$-values for $H_{s}^{\gamma}$ from Section$~\ref{sec:pvaluecombinations}$ in simulations. The marginal $p$-values $p_{1},\ldots,p_{s}$ in our simulations are given by two different models. 

\subsection{Models for \texorpdfstring{$p$}{p}-value generation}
\label{subsec:pvaluemodels}

We consider Beta-distributed $p$-values, which has also been used for example by \cite{loughin2004systematic}. For a parameter value $\boldsymbol{\theta}\in\Theta=\mathbb{R}^{s}$, we define the density function $f_{\theta_{i}}$ of the $i$-th $p$-value as  

\begin{equation*}
\begin{cases}
f_{\theta_{i}}\sim\mathrm{Beta}\left(1-\theta_{i},1\right), & \theta_{i}\leq0,\\
f_{\theta_{i}}\sim\mathrm{Beta}\left(1,1+\theta_{i}\right), & \theta_{i}>0,
\end{cases}
\end{equation*}
where $\mathrm{Beta}(\alpha,\beta)$ denotes the Beta-distribution with parameters $\alpha$ and $\beta$. 

As the second model, we consider the Normal-Model, where the $p$-values result from a Gaussian shift model with known variance $\sigma^{2}>0$. Here, we define the density function of the $i$-th $p$-value as
\begin{equation*}
f_{\theta_{i}}(t)=\frac{\varphi_{(\theta_{i},\sigma^{2})}\left(\Phi_{(0,\sigma^{2})}^{-1}(1-t)\right)}{\varphi_{(0,\sigma^{2})}\left(\Phi_{(0,\sigma^{2})}^{-1}(1-t)\right)},\;t\in[0,1],
\end{equation*}
where $\varphi_{(\mu,\sigma^{2})}$ is the density function, and $\Phi_{(\mu,\sigma^{2})}^{-1}$ the quantile function of the normal distribution with expected value $\mu$ and variance $\sigma^{2}$.

\begin{lemma}
\label{lm:assumptions}
Both models satisfy Assumptions $(A1) - (A3)$  from Section~\ref{sec:modelsetup}.
\end{lemma}
\textit{Proof}: Assumption $(A1)$ is clear. Regarding assumption $(A2)$, the Beta-distribution $\mathrm{Beta}(1,1)$ with parameters $\alpha=\beta=1$ is the $\mathrm{Uni}[0,1]$ distribution, and therefore $f_{0}=\textbf{1}_{[0,1]}(t)$. Analogously this is also the case for the Normal-Model. 

For assumption $(A3)$, we analyze the cdf of the $i$-th $p$-value. In the Beta-Model, if $\theta_{i}\leq 0$, the $p$-value $p_{i}(\boldsymbol{X})$ is $\mathrm{Beta}(1-\theta_{i},1)$-distributed with cdf $F_{\theta_{i}}(t)=t^{-\theta_{i}+1}\textbf{1}_{[0,1]}(t)+\textbf{1}_{(1,\infty)}(t)$, which is decreasing for decreasing $\theta_{i}$ and each fixed $t$. If $\theta_{i}>0$, the cdf of $p_{i}(\boldsymbol{X})$ is $F_{\theta_{i}}(t)=\left(1-(1-t)^{\theta_{i}+1}\right)\textbf{1}_{[0,1]}(t)+\textbf{1}_{(1,\infty)}(t)$, which is increasing in $\theta_{i}$ and each fixed $t$. Thus assumption $(A3)$ is fulfilled in the Beta-Model.

In the Normal-Model, we refer to Remark$~\ref{rm:likelihood}$, where the test statistic $T_{i}(\boldsymbol{X})$ is normally distributed with expected value $\theta_{i}$ and (known) variance $\sigma^{2}$.

This concludes the proof of Lemma~$\ref{lm:assumptions}$.

In our simulations below, we draw the true parameter value $\theta_{i}$ uniformly from intervals $\left[\theta^{b}_{i},0\right]$ and $\left(0,\theta^{b}_{i}\right]$ if $\theta^{b}_{i}\leq 0$ or $\theta^{b}_{i}>0$, respectively. Similarly to \cite{loughin2004systematic}, we write $\theta^{b}_{i}=r\mu^{b}_{i}$, $r>0$, $i=1,\ldots,s$. Holding each $\mu^{b}_{i}$ constant, we can vary the potential ``signal strength'' of each $p$-value with $r$, i.e. with increasing $r$ the $i$-th $p$-value $p_{i}$ gets stochastically larger / more conservative under $H_{i}$ (assuming $\mu^{b}_{i}\neq0$) and stochastically smaller under $K_{i}$ under $\theta^{b}_{i}$. 

Table$~\ref{tab:evidencestructures}$ summarizes the different patterns $(\mu_{1}^{b},\ldots,\mu_{s}^{b})^{T}$ that we use in our simulations, cf. also Table$~3$ in \cite{loughin2004systematic}. We set the number of studies to $s=6$. The patterns are first ordered in their amount of false null hypotheses, i.e. the amount of indices $i$ with $\mu^{b}_{i}>0$. Patterns with the same amount of false null hypotheses are then ordered decreasingly in their order of dispersion $\sum_{i}(\mu^{b}_{i})^{2}$. Furthermore, we denote by pattern $j$c the conservative version of pattern $j$, where we replace each $\mu_{i}^{b}=0$ by $\mu_{i}^{b}=-2$. Patterns $10$ -- $13$ have no conservative versions.

\begin{table}
\begin{centering}
\begin{tabular}{c| c c c c c c |c}
 
Pattern & $\mu^{b}_{1}$ & $\mu^{b}_{2}$ & $\mu^{b}_{3}$ & $\mu^{b}_{4}$ & $\mu^{b}_{5}$ & $\mu^{b}_{6}$ & $\sum_{i}(\mu^{b}_{i})^{2}$\tabularnewline
\hline 

1 & 0 & 0 & 0 & 0 & 1 & 5 & 26\tabularnewline

2 & 0 & 0 & 0 & 0 & 3 & 3 & 18\tabularnewline

3 & 0 & 0 & 0 & 1 & 1 & 4 & 18\tabularnewline

4 & 0 & 0 & 0 & 2 & 2 & 2 & 12\tabularnewline
 
5 & 0 & 0 & 1 & 1 & 1 & 3 & 12\tabularnewline

6 & 0 & 0 & 1.5 & 1.5 & 1.5 & 1.5 & 9\tabularnewline

7 & 0 & 0.5 & 0.5 & 0.5 & 0.5 & 4 & 17\tabularnewline

8 & 0 & 1 & 1 & 1 & 1 & 2 & 8\tabularnewline

9 & 0 & 1.2 & 1.2 & 1.2 & 1.2 & 1.2 & 7.2\tabularnewline

10 & 0.2 & 0.2 & 0.2 & 0.2 & 0.2 & 5 & 25.2\tabularnewline

11 & 0.5 & 0.5 & 0.5 & 0.5 & 2 & 2 & 9\tabularnewline

12 & 0.5 & 0.5 & 1.25 & 1.25 & 1.25 & 1.25 & 6.75\tabularnewline
 
13 & 1 & 1 & 1 & 1 & 1 & 1 & 6\tabularnewline

\end{tabular}
\par\end{centering}
\caption{The evidence patterns with uniformly distributed $p$-values under nulls}
\label{tab:evidencestructures}
\end{table}

\subsection{Calculation of Bayes factors}
\label{subsec:bayesfactors}

We calculate the marginal Bayes factors for the two approaches that utilize $e$-value combinations under the same models as in Section$~\ref{subsec:pvaluemodels}$. For this, we need to make some assumptions about the prior distributions of the parameter values under the null hypotheses and under the alternatives.

We assume that it is known beforehand whether the marginal null hypotheses $H_{j}$, $j=1,\ldots,s$, are simple (Patterns$~1$ -- $13$) or composite (Patterns$~1c$ -- $9c$). In both cases we calculate the Bayes factors under the assumption that all parameter values $\boldsymbol{\theta}=(\theta_{1},\ldots,\theta_{s})^{T}\in K_{i}$ are such that the $i$-th component $\theta_{i}$ is drawn uniformly from the interval $(0,5r]$. Under simple null hypotheses the resulting Bayes factors are $e$-values, i.e. they have expected values not larger than one, cf. \cite{vovk2020combining}. Under composite null hypotheses, we calculate the Bayes factors under the assumption that $\theta_{i}$ is uniformly distributed on $[-3r,0]$ if $\boldsymbol{\theta}\in H_{i}$. The numbers $5$ and $-3$ were chosen such that the true underlying parameter values $\theta_{i}$ drawn from any of the patterns in Table$~\ref{tab:evidencestructures}$ are included in $(0,5r]$ or $[-3r,0]$.

In the latter case, the resulting Bayes factors are not valid $e$-values for the marginal null hypotheses, i.e. their expected value is larger than one for some parameters under the null. More specifically, the $i$-th Bayes factor has an increasing expected value under increasing $\theta_{i}\in[-3r,0]$. Therefore, under all parameters $\boldsymbol{\theta}\in H_{i}$, it has its largest expected value when $\theta_{i}=0$. See Appendix for a proof of this. To create valid $e$-values we therefore divide the Bayes factors in Patterns$~1c$ -- $9c$ by this expected value. Note, that computing this constant requires no extra information beyond the information necessary for calculating the Bayes factors.

\subsection{Power Simulations}
\label{subsec:powersimulations}

The power of a $p$-value $p$ under a parameter value $\boldsymbol{\theta}$ in the alternative given a significance level $\alpha\in(0,1)$ is defined as $\mathbb{P}_{\boldsymbol{\theta}}(p\leq\alpha)$. Under various parameter settings, where $H_{s}^{\gamma}$ is false, we approximate the relative power (relative to the best performing one in each setting, where we set the significance level to $\alpha=0.05$) of each $p$-value combination via a Monte-Carlo simulation with $100{,}000$ repetitions.

First, we look at different evidence structures in Table$~\ref{tab:evidencestructures}$. For a pattern where $H_{s}^{\gamma}$ is false, the evidence for its rejection can be focused in few false $p$-values or it can be more evenly spread between the false $p$-values, compare for example Pattern$~3$ versus Pattern$~4$. Furthermore, we want to investigate how the choice of $\gamma$ affects the performance of the $p$-value combination functions for different types of evidence structures.

\subsubsection{Evidence Structures}
\label{subsubsec:evidencestructures}

For the sake of clearness of the graphical displays, we decided to only display the simulation results for the Stouffer, Fisher and minimum $p$-value as well as the product of the $e$-values (called $e$-product). The harmonic mean and the arithmetic mean of the $e$-values (not displayed) performed badly to mediocrely throughout.

\begin{figure}
\centering{}\includegraphics[scale=0.5]{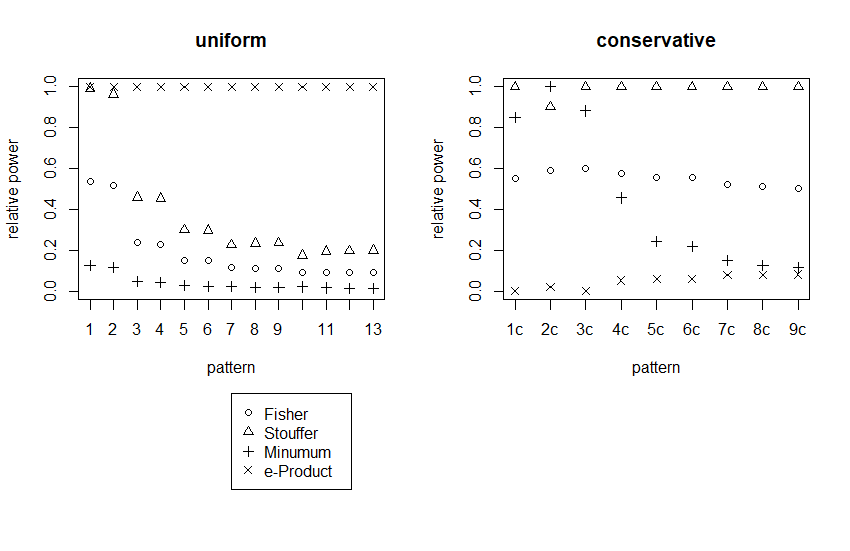}\caption{Relative power of the combined $p$-values under the Beta-Model with $\gamma=2$ and $r=1$.}
\label{fig:beta1}
\end{figure}

\begin{figure}
\centering{}\includegraphics[scale=0.5]{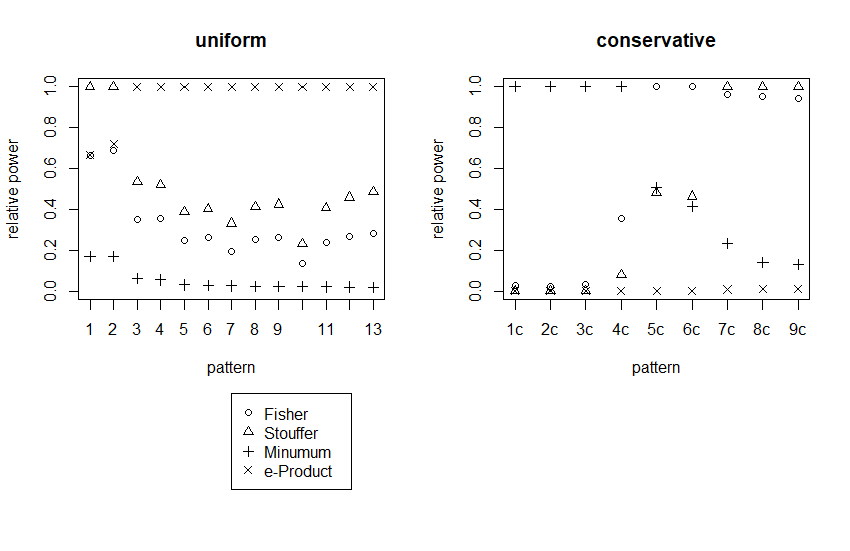}\caption{Relative power of the combined $p$-values under the Beta-Model with $\gamma=2$ and $r=5$.}
\label{fig:beta2}
\end{figure}

\begin{figure}
\begin{centering}
\includegraphics[scale=0.5]{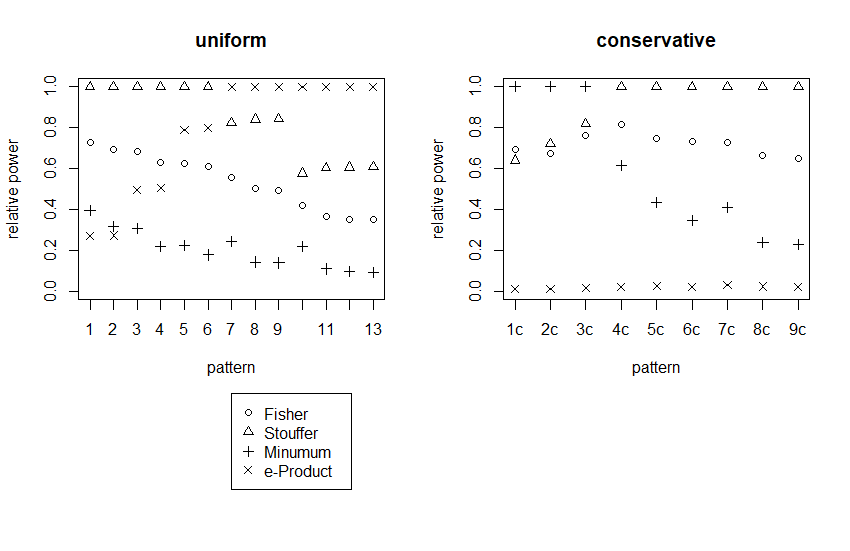}
\par\end{centering}
\centering{}\caption{Relative power of the combined $p$-values under the Normal Model with $\gamma=2$ and $r=0.5\sigma$.}
\label{fig:normal1}
\end{figure}

\begin{figure}
\begin{centering}
\includegraphics[scale=0.5]{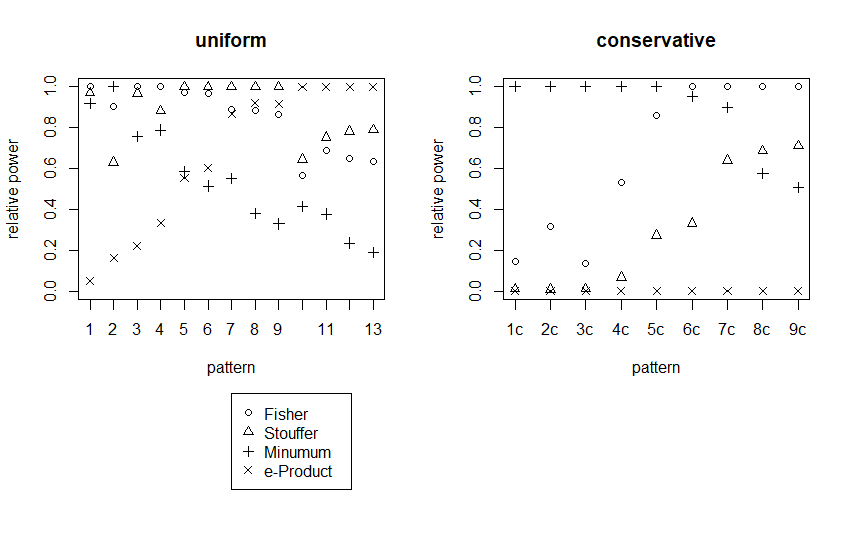}
\par\end{centering}
\centering{}\caption{Relative power of the combined $p$-values under the Normal Model with $\gamma=2$ and $r=1.5\sigma$.}
\label{fig:normal2}
\end{figure}

Figures$~\ref{fig:beta1}$ -- $\ref{fig:beta2}$ have been derived under the Beta-Model of generating the marginal $p$-values. We set $\gamma=2$ and the significance strengths to $r=1,5$ in Figures$~\ref{fig:beta1}$ and $\ref{fig:beta2}$, respectively. In Figures$~\ref{fig:normal1}$ -- $\ref{fig:normal2}$, we generated the marginal $p$-values according to the Normal-Model with $\sigma=1/\sqrt{50}$. We set $\gamma=2$ and the significance strengths in the figures to $r=0.5\sigma,1.5\sigma$, respectively. The distribution of the $p$-values under nulls is indicated above the graphics. 

At first, we summarize the observations of the two figures with lower signal strength $r$, Figures$~\ref{fig:beta1}$ and $\ref{fig:normal1}$. If the null $p$-values are $\mathrm{Uni}[0,1]$, the Stouffer $p$-value has the highest power to reject $H_{6}^{2}$ if the evidence is more focused (lower pattern number). If the evidence is more spread out the $e$-product has the highest power. The power of the Fisher $p$-value is slightly below that of the Stouffer $p$-value and the minimum $p$-value performs badly.

If the null $p$-values are conservative in Figures$~\ref{fig:beta1}$ and $\ref{fig:normal1}$, the minimum $p$-value performs best if the evidence is focused. If the evidence is more spread out, both the Stouffer and, to a lesser extent, the Fisher $p$-value have the highest power. The $e$-product performs badly in this case.

In Figures$~\ref{fig:beta2}$ and $\ref{fig:normal2}$ we used a higher signal strength $r$. If the null hypotheses are simple and the evidence is focused, the Stouffer $p$-value has the highest power under the Beta-Model. If the evidence is more spread out the $e$-product has the highest power. Under the Normal-Model the minimum $p$-value and the Fisher $p$-value are most powerful if the evidence is focused and the Stouffer $p$-value if the evidence is less focused. The $e$-product is most powerful if the evidence is spread out. 

If the null $p$-values are conservative in Figures$~\ref{fig:beta2}$ and $\ref{fig:normal2}$, the minimum $p$-value has the highest powers in the first patterns. The Stouffer has highest power under the Beta-Model and the Fisher $p$-value has highest power under the Normal-Model in the latter half of the patterns. The $e$-product performs badly throughout all the patterns.

To summarize, the $e$-product work best if we consider the non-conservative versions of the patterns, especially if the null $p$-values are uniformly distributed and the evidence is spread. If the null $p$-values are uniformly distributed and the evidence is focused the Stouffer $p$-value has the highest power. In the conservative patterns, the minimum $p$-value works best for lower pattern numbers. For the higher pattern numbers the Stouffer $p$-value works well if the signal strength is lower, and the Fisher $p$-value works well if the signal strength is higher. The results between the two $p$-value generating models are mostly similar.

\subsubsection{The parameter \texorpdfstring{$\gamma$}{gamma}}
\label{subsubsec:gamma}

In this section, we investigate the influence of $\gamma$ on the relative performances of the $p$-value combination functions. More specifically, we chose Patterns$~7$ and $7c$, in which five of the null hypotheses are false, and thus $H_{s}^{\gamma}$ is false for $\gamma=1,\ldots,5$. Again, we look at the relative powers of the $p$-value combination functions, relative to the best performing combination function in each setting.

\begin{figure}
\begin{centering}
\includegraphics[scale=0.5]{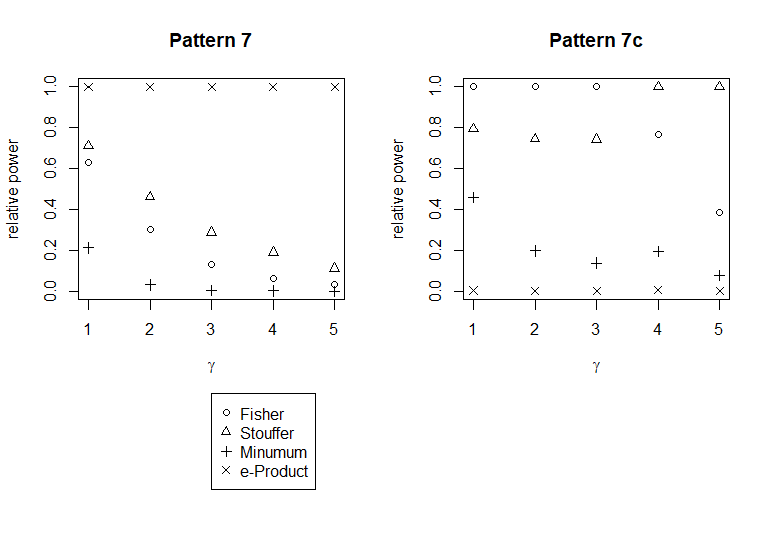}
\par\end{centering}
\centering{}\caption{Relative power of the combined $p$-values under the Beta-Model in Patterns$~7$ and $7c$, with $\gamma=1,\ldots,5$ and $r=10$.}
\label{fig:betagamma}
\end{figure}

In Figure$~\ref{fig:betagamma}$ we employed the Beta-Model. The $p$-value obtained from the $e$-product has the highest power throughout all values of $\gamma=1,\ldots,5$ in Pattern$~7$. With increasing $\gamma$, the power of the other combined $p$-values fall off faster than the power of the $e$-product. Under Pattern$~7c$ the Fisher and the Stouffer $p$-value have the highest power, the former if $\gamma=1,2,3$ and the latter if $\gamma=4,5$. The $e$-value approach, which is adjusted in this case, performs much worse.

\begin{figure}
\begin{centering}
\includegraphics[scale=0.5]{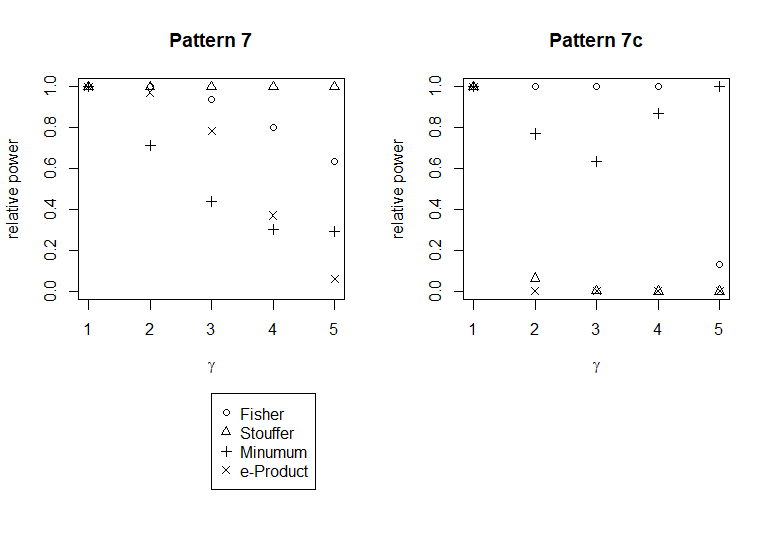}
\par\end{centering}
\centering{}\caption{Relative power of the combined $p$-values under the Normal-Model in Patterns$~7$ and $7c$, with $\gamma=1,\ldots,5$ and $r=3\sigma$.}
\label{fig:normalgamma}
\end{figure}

In Figure$~\ref{fig:normalgamma}$ we used the Normal-Model. The $p$-values are close in power for $\gamma=1$, their power is essentially $1$ in absolute values. For $\gamma>1$ in Pattern$~7$, the power of all the $p$-values fall relative to the power of the Stouffer $p$-value. The Fisher $p$-value performs relatively well and its power only falls off after $\gamma=3$. In Pattern$~7c$, the Fisher $p$-value has the highest power if $\gamma$ is between $2$ and $4$. For $\gamma=5$, the minimum $p$-value has the highest power.

While the results under the Beta-Model suggest the superiority of the approach using $e$-values in Pattern$~7$, the results under the Normal-Model are more diverse. In both models, the Fisher $p$-value has higher power than the Stouffer $p$-value if the null $p$-values are conservative, and vice versa if the null $p$-values are uniformly distributed. Furthermore, the minimum $p$-value works (relatively) well if $\gamma$ is large, especially if $\gamma$ is the true number of false null hypotheses, which is five in Patterns$~7$ and $7c$. 

We illustrate this with a short example under the assumption that $H_{s}^{\gamma}$ is false, that is, at least $\gamma$ of the null hypotheses $H_{1},\ldots,H_{s}$ are false. In terms of power, the worst case scenario for a monotonic combination function occurs if the $p$-values are as large as possible, which is the case if $\gamma$ null hypotheses are false with corresponding $p$-values that are uniformly distributed, and $s-\gamma$ true null hypotheses with corresponding $p$-values that are almost surely $1$. Note, that the distribution of false $p$-values is lower bounded by $\mathrm{Uni}[0,1]$ due to Assumption$~(A3)$. Under this worst case scenario, the ordered, marginal $p$-values are $U_{(1)},\ldots,U_{(\gamma)},1,\ldots,1$, therefore the $s-\gamma+1$ largest $p$-values are $U_{(\gamma)},1,\ldots,1$. Thus, testing for $H_{s}^{\gamma}$, the minimum $p$-value only directly evaluates $U_{(\gamma)}$, while averaging methods for instance by Fisher and Stouffer evaluate $U_{(\gamma)},1,\ldots,1$, in this extreme case. Testing for $H_{s}^{\gamma-1}$ (which is also false if $H_{s}^{\gamma}$ is false), the minimum $p$-value evaluates $U_{(\gamma-1)}$ whereas Fisher and Stouffer now consider $U_{(\gamma-1)},U_{(\gamma)},1,\ldots,1$. The ratio of non-one to one $p$-values increases with decreasing $\gamma$, which favors averaging methods more than the minimum $p$-value.

\subsection{Null \texorpdfstring{$p$}{p}-value simulations}
\label{subsec:nullsimulations}

In the previous simulations we only considered the case of false null hypotheses $H_{s}^{\gamma}$. In this section we investigate the behavior of the $p$-value combination functions under the null hypothesis $H_{s}^{\gamma}$.

Each of the presented $p$-value combination functions in Section$~\ref{sec:pvaluecombinations}$ is valid for the null hypothesis $H_{s}^{\gamma}$, i.e. they are stochastically at least as large as $\mathrm{Uni}[0,1]$. Conservative $p$-values, that is, $p$-values that are stochastically larger than $\mathrm{Uni}[0,1]$, are common under composite null hypotheses, where the $p$-value is only calibrated with respect to the LFC parameter under the null. While still maintaining the type I error control, conservative null $p$-values can be problematic in several multiple testing setups that require uniformly distributed null $p$-values. 

In simulations, we approximate the cdf $F_{\boldsymbol{\theta}}(\alpha)$ at point $\alpha$ of the combined $p$-values for $H_{s}^{\gamma}$ under different parameter values $\boldsymbol{\theta}$ for which the null hypothesis $H_{s}^{\gamma}$ hold. If $\alpha$ is the significance level, the value $F_{\boldsymbol{\theta}}(\alpha)$ for such $\boldsymbol{\theta}$ is the probability of a false rejection of $H_{s}^{\gamma}$. Since we only consider valid $p$-values, it holds $F_{\boldsymbol{\theta}}(\alpha)=\alpha$ if the $p$-value is $\mathrm{Uni}[0,1]$-distributed and $F_{\boldsymbol{\theta}}(\alpha)\leq\alpha$ if the $p$-value is conservative. It is of interest that $F_{\boldsymbol{\theta}}(\alpha)$ is as close to $\alpha$ as possible. One example is the problem of estimating the proportion $\pi_{0}$ of true null hypotheses in a multiple testing setup with the Schweder-Spj{\o}tvoll estimator $\hat{\pi}_{0}(\alpha)$, cf. \cite{schweder1982plots}. The estimator $\hat{\pi}_{0}(\alpha)$ utilizes the marginal $p$-values, and its bias $\mathbb{E}_{\boldsymbol{\theta}}\left[\hat{\pi}_{0}(\alpha)\right]-\pi_{0}\geq 0$ increases with decreasing $F_{\boldsymbol{\theta}}(\alpha)$ for any of the marginal $p$-values, cf. \cite{hoang2021randomized}. 

The choice $\lambda=\alpha$ in the Schweder-Spj{\o}tvoll estimator $\hat{\pi}_{0}(\lambda)$ was proposed by \cite{blanchard2009adaptive}. For arbitrary parameter values $\lambda\in[0,1)$ in the Schweder-Spj{\o}tvoll estimator we have to look at the entire cdf $F_{\boldsymbol{\theta}}$. If the $p$-value is $\mathrm{Uni}[0,1]$-distributed, its cdf is a straight line between $(0,0)$ and $(1,1)$, and more conservative $p$-values have a cdf below that line. For select parameters values $\boldsymbol{\theta}$, we approximate the cdf of some of the $p$-values.

\begin{figure}
\begin{centering}
\includegraphics[scale=0.5]{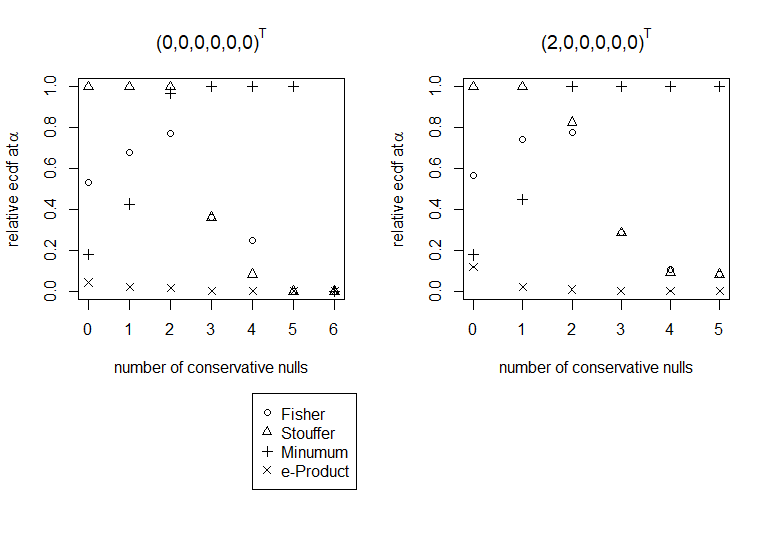}
\par\end{centering}
\centering{}\caption{The $p$-values are generated under the Beta-Model. The graphs display the approximations of the cdf at $\alpha$, relative to the (average) maximum estimation in the respective simulation, of the $p$-value combination functions testing the null hypothesis $H_{6}^{2}$, via Monte Carlo simulation with $100{,}000$ repetitions. The underlying pattern is indicated above the graphs, the signal strength $r$ is $5$. We replace $0$ by $-1$ if conservative, the number of times we do this varies on the horizontal axis.}
\label{fig:falserejections1}
\end{figure}

\begin{figure}
\begin{centering}
\includegraphics[scale=0.5]{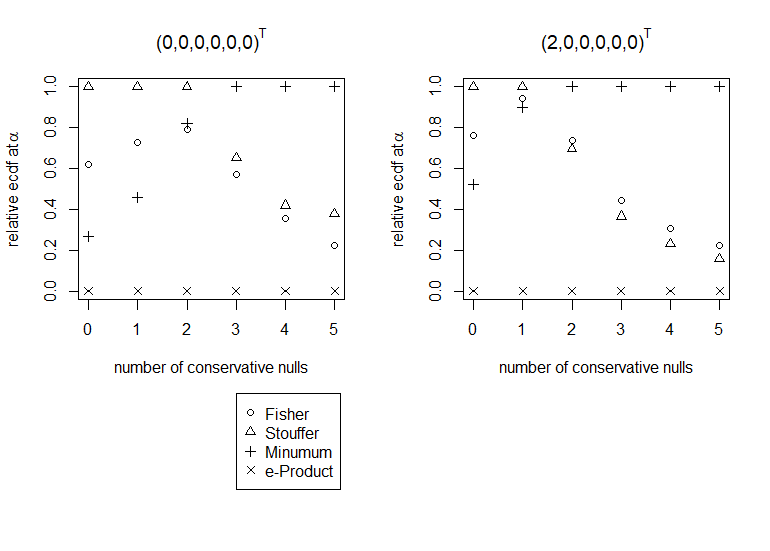}
\par\end{centering}
\centering{}\caption{The $p$-values are generated under the Normal-Model with standard deviation $\sigma=1/\sqrt{50}$. The graphs display the approximations of the cdf at $\alpha$, relative to the (average) maximum estimation in the respective simulation, of the $p$-value combination functions testing the null hypothesis $H_{6}^{2}$, via Monte Carlo simulation with $100{,}000$ repetitions. The underlying pattern is indicated above the graphs, the signal strength $r$ is $1.5\sigma$. We replace $0$ by $-1$ if conservative, the number of times we do this varies on the horizontal axis.}
\label{fig:falserejections2}
\end{figure}

Figures$~\ref{fig:falserejections1}$ and $\ref{fig:falserejections2}$ plot the empirical cumulative distribution functions (ecdfs) of the $p$-value combinations at point $\alpha$, relative to the largest one in each setting, generated by a Monte-Carlo simulation with $100{,}000$ repetitions, where we test for the rejection of $H_{6}^{2}$, i.e. that at least two null hypotheses are false. We use $(\mu_{1}^{b},\ldots,\mu_{6}^{b})^{T}=(0,\ldots,0)^{T}$ on the left and $(\mu_{1}^{b},\ldots,\mu_{6}^{b})^{T}=(2,0,\ldots,0)^{T}$ on the right graphs. Furthermore, we replace $0$ by $-1$ in $(\mu_{1}^{b},\ldots,\mu_{6}^{b})^{T}$ if the respective null is conservative. The number of times we do this is indicated on the horizontal axis.

In Figure$~\ref{fig:falserejections1}$ we used the Beta-Model and in Figure$~\ref{fig:falserejections2}$ we used the Normal-Model. The results are similar. The Stouffer $p$-value has the highest ecdf at $\alpha$ if the number of conservative nulls is low (below two or three), the minimum $p$-value has the highest ecdf at $\alpha$ if that number is higher. The Fisher $p$-value has mediocre performances and comes closer to the best $p$-values on the right graphs. The $e$-product has the lowest ecdf values at $\alpha$.

Additionally, we display the ecdfs of the Stouffer, the Fisher and the minimum $p$-value in the cases of $1$ and $4$ conservative null $p$-values under the Beta-Model like in the right plot of Figure$~\ref{fig:falserejections1}$. The values for $(\mu_{1}^{b},\ldots,\mu_{6}^{b})^{T}$ are displayed above the plots, $r$ is $5$. First, we notice that the ecdfs are closer to the identical line on the left plot than they are on the right. On the left plot the ecdfs are close to each other, whereas on the right one the ecdf of the minimum $p$-value is noticeably closer to the identical line compared to the other two ecdfs. Another difference is that the ecdfs are not ordered consistently at each point $t\in[0,1]$ on the left plot, which implies that the corresponding $p$-values are not stochastically ordered. On the right plot, however, the ecdf of the minimum $p$-value seems to be the largest at each point $t\in[0,1]$, and therefore the minimum $p$-value is stochastically closest to $\mathrm{Uni}[0,1]$ in this more conservative setting.

\begin{figure}
\begin{centering}
\includegraphics[scale=0.5]{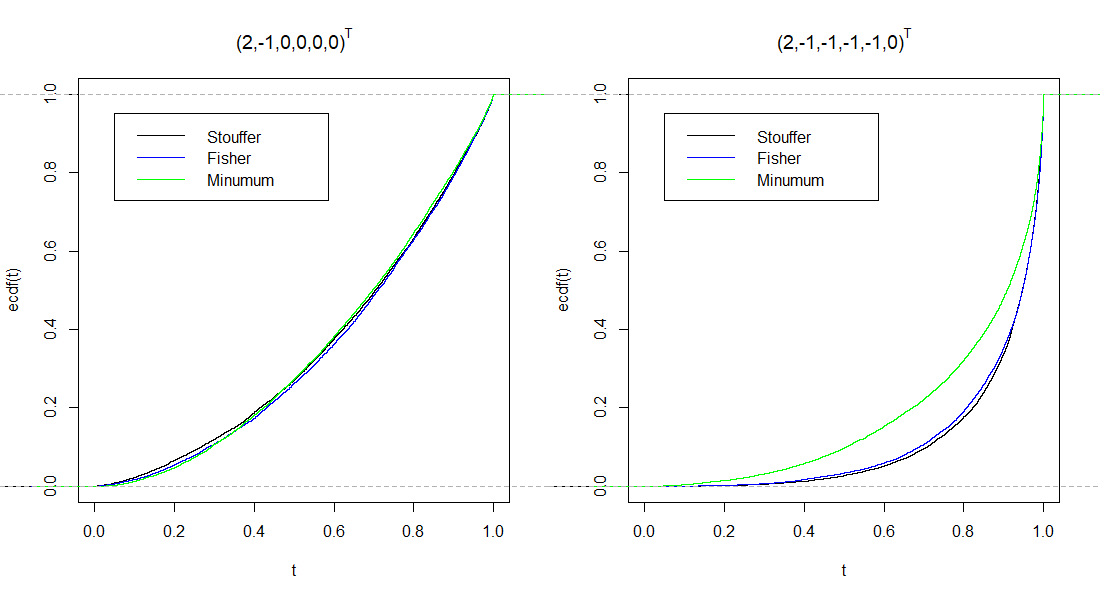}
\par\end{centering}
\centering{}\caption{The ecdfs from a Monte-Carlo simulation with $10{,}000$ repetitions of the Fisher, Stouffer and minimum $p$-values under the parameter $(\mu_{1}^{b},\ldots,\mu_{6}^{b})^{T}=(2,-1,0,\ldots,0)^{T}$ on the left side and  $(\mu_{1}^{b},\ldots,\mu_{6}^{b})^{T}=(2,-1,\ldots,-1,0)^{T}$ in $H_{6}^{2}$. The marginal $p$-values have been generated with the Beta-Model with signal strength $r=5$.}
\label{fig:ecdfs}
\end{figure}

\section{Discussion}
\label{sec:discussion}

We compared a number of $p$-value combination functions for independent $p$-values and compared their power under the alternative hypothesis and their degree of conservativity under the null hypothesis when testing the partial conjunction hypothesis $H_{s}^{\gamma}$. Among the $p$-value combination functions that we considered in this paper, we can roughly distinguish between two classes. One are the $p$-values that rely on a weighted average of the $s-\gamma+1$ largest $p$-values and one that only evaluates the ($s-\gamma+1$)-th largest $p$-value. They mainly differ in how they deal with spread out evidence versus focused evidence, and conservative versus uniformly distributed null $p$-values. 

Among the considered $p$-value combination functions, the approaches that utilize Bayes factors as $e$-values, work best if the null hypotheses $H_{1},\ldots,H_{s}$ are simple, and the Stouffer, the Fisher and the minimum $p$-values have the best results, if the null hypotheses are composite. The three latter $p$-values excel in different situations under the alternative and the null hypothesis. Under the alternative, the Stouffer $p$-value appears to have the highest power if the marginal $p$-values are uniformly distributed and the signal strength is low. The Fisher $p$-value works better than the other combination functions if the marginal null $p$-values are conservative and the evidence is spread out between several false null hypotheses. The minimum method works best if the evidence is focused on few false null hypotheses, especially when the null $p$-values are conservative. Under the null hypothesis $H_{s}^{\gamma}$, the Stouffer $p$-value is closest to uniformity if the marginal null $p$-values are $\mathrm{Uni}[0,1]$-distributed. However, if there is at least one conservative null $p$-value, the minimum method works better than the other combination methods. The Fisher $p$-value is stochastically closer to $\mathrm{Uni}[0,1]$ than the Stouffer $p$-value if some of the marginal null $p$-values are conservative. One major difference between the Stouffer and the Fisher $p$-value is that the latter emphasizes the smallest $p$-values more and is thus less affected by conservative $p$-values, cf. \cite{owen2009karl}, which coincides with our results. While our selection of combination functions for $H_{s}^{\gamma}$ is not exhaustive, the overall conclusions in this paper can be generalized to $p$-values that evaluate an average of all $p$-values versus $p$-values that place more weight on the smallest $p$-values.

Since this paper is concerned about replicability analyses, we limited our research to $p$-values from independent tests. It is interesting to see how the results differ, if the $p$-values are dependent. Methods that are designed for combining independent $p$-values, like the Fisher, Stouffer and minimum $p$-value, tend to exaggerate the evidence for the the alternative and the null hypothesis if the marginal $p$-values are positively correlated, cf. \cite{alves2014accuracy}. Introducing weights can be helpful in this matter. Furthermore, we only consider two models for the generation of the marginal $p$-values. The Beta-Model was also used in \cite{loughin2004systematic} and our results are similar when $\gamma=1$. The Normal-Model had similar results as well. In both models the $p$-values have non-decreasing densities under the null hypothesis and non-increasing under the alternative. Deviation from this assumption is not considered here and is an attractive topic for future research.

\section*{Appendix}
\label{sec:appendix}

\subsection*{Some results regarding \texorpdfstring{$e$}{e}-variables and Bayes factors in our models}

The following lemma is analogous to Lemma $1$ in \cite{benjamini2008screening}, for $e$-variables instead of $p$-values. 

\begin{lemma*}
\label{lm:eval}

Let $e_{1},\ldots,e_{s}$ be valid $e$-variables for $H_{1},\ldots,H_{s}$, respectively. If $h_{0}:[0,\infty]^{s-\gamma+1}\rightarrow[0,\infty]$ is a valid, symmetric and monotonically non-decreasing $e$-variable combination function for $H_{s-\gamma+1}^{1}$, then 
\begin{equation*}
h\left(e_{1},\ldots,e_{s}\right)=h_{0}\left(e_{(1)},\ldots,e_{(s-\gamma+1)}\right)
\end{equation*}
is a valid $e$-variable for $H_{s}^{\gamma}$.

\end{lemma*}
\begin{proof}

We have to show that 
\begin{equation}
\label{eq:valide}
\mathbb{E}_{\theta}\left[h\left(e_{1},\ldots,e_{s}\right)\right]\leq 1
\end{equation} 
holds for all $\theta\in H_{s}^{\gamma}$. We determine the parameters $\theta\in H_{s}^{\gamma}$ for which $\mathbb{E}_{\theta}\left[h\left(e_{1},\ldots,e_{s}\right)\right]$ is largest and show that \eqref{eq:valide} holds. Since $h_{0}$ is monotonically non-decreasing in all its arguments, so is $h$. Under $H_{s}^{\gamma}$, where at least $s-\gamma+1$ of the null hypotheses $H_{1},\ldots,H_{s}$ are true, the worst case for \eqref{eq:valide} occurs, when exactly $s-\gamma+1$ of the null hypotheses $H_{1},\ldots,H_{s}$ are true, since $e$-variables are unbounded under alternatives. The $\gamma-1$ $e$-variables that correspond to the false null hypotheses are as large as possible in the worst case, for simplicity they are $\infty$. 

Without loss of generality, assume that the null hypotheses $H_{1},\ldots,H_{s-\gamma+1}$ are the true ones. The ordered $e$-values 
\[(e_{(1)},\ldots,e_{(s)})=(\tau(e_{1},\ldots,e_{s-\gamma+1}),\infty,\ldots,\infty), 
\]
where $\tau$ is a permutation map, are such that the $s-\gamma+1$ smallest $e$-variables correspond to the true null hypotheses. Therefore, in the worst case, the combination of these $e$-variables 
\begin{equation*}
h\left(e_{(1)},\ldots,e_{(s-\gamma+1)},\infty,\ldots,\infty\right)=h_{0}\left(e_{1},\ldots,e_{s-\gamma+1}\right)
\end{equation*}  
has an expected value not greater than $1$, according to the definition of $h_{0}$.

\end{proof}

The next result helps us construct valid $e$-variables under composite null hypotheses. We assume that a $p$-value model as in Section$~\ref{sec:modelsetup}$ is given. Furthermore, we assume that for all $i$ the Lebesgue density $f_{\theta_{i}}$ of $p_{i}$ is monotonically decreasing if $\theta_{i}\leq 0$ and monotonically increasing if $\theta_{i}>0$. This latter assumption is for example fulfilled under the conditions in Remark$~\ref{rm:likelihood}$. To see this, we use Assumption$~(A2)$ and note that if the distributions $(p_{i}(\boldsymbol{X})^{(-\theta_{i})})_{\theta_{i}}$ are likelihood ratio ordered, then $f_{\theta_{i}}(t)/f_{\tilde{\theta}_{i}}(t)$ is non-decreasing in $t$, if $\theta_{i}\leq\tilde{\theta}_{i}$. 

Let $i$ be fixed. For given Bayes marginal probability distributions $\mathbb{P}_{0}$ under $H_{i}$ and $\mathbb{P}_{1}$ under $K_{i}$ for the parameter values, we define the Bayes factor as

\begin{equation}
\label{eq:BF}
\mathrm{BF}(p):=\frac{\int f_{\theta_{i}}(p)d\mathbb{P}_{1}(\theta_{i})}{\int f_{\theta_{i}}(p)d\mathbb{P}_{0}(\theta_{i})}.
\end{equation}

Its expected value under $\theta_{i}$ is 
\begin{equation*}
\mathbb{E}_{\theta_{i}}\left[\mathrm{BF}\right]=\int\mathrm{BF}(p)f_{\theta_{i}}(p)dp,
\end{equation*}
which is required to not be larger than $1$ under each $\theta_{i}\in H_{i}$ for $\mathrm{BF}$ to be a valid $e$-variable for $H_{i}$. The following result helps determine whether $\mathrm{BF}$ is an $e$-variable.

\begin{lemma*}

Under the assumptions from above and for the Bayes factor $\mathrm{BF}$ as in \eqref{eq:BF}, the expected value $\mathbb{E}_{\theta_{i}}\left[\mathrm{BF}\right]$ of $\mathrm{BF}$ is non-decreasing in $\theta_{i}$. 
 
\end{lemma*}

\begin{proof}

With the assumptions we made, the Bayes factor $\mathrm{BF}(p)$ is non-decreasing in $p$. Thus the expected value of (the distribution) $\left[\mathrm{BF}(p_{i}(\boldsymbol{X}))\right]^{(\theta_{i})}$ decreases with stochastically increasing $p_{i}(\boldsymbol{X})^{(\theta_{i})}$ and therefore with decreasing $\theta_{i}$.

\end{proof}

Both the Beta-Model and the Normal-Model fulfill the assumptions for this lemma. With this lemma, the Bayes factor $\mathrm{BF}$ has its largest expected value under $H_{i}$ under $\theta_{i}=0$, regardless of the priors. Therefore, $\mathrm{BF}/\mathbb{E}_{0}[\mathrm{BF}]$ is an $e$-variable for $H_{i}$ with expected value $1$ under $\theta_{i}=0$. 

Under the same assumptions, it is also easy to see that the Bayes factor 
\begin{equation*}
\mathrm{BF}_{0}(p):=\frac{\int f_{\tilde{\theta}_{i}}(p)d\mathbb{P}_{1}(\theta_{i})}{f_{0}(p)}=\int f_{\tilde{\theta}_{i}}(p)d\mathbb{P}_{1}(\theta_{i})
\end{equation*}
that assumes a simple null hypothesis is an $e$-variable with expected value $1$ under $\theta_{i}=0$ even if $H_{i}$ is really composite.

\bibliographystyle{apalike}
\bibliography{Combi-replicability-arXiv}

\begin{thebibliography}{}

\bibitem[Alves and Yu, 2014]{alves2014accuracy}
Alves, G. and Yu, Y.-K. (2014).
\newblock Accuracy evaluation of the unified p-value from combining correlated
  p-values.
\newblock {\em PloS ONE}, 9(3):e91225.

\bibitem[Benjamini and Heller, 2008]{benjamini2008screening}
Benjamini, Y. and Heller, R. (2008).
\newblock Screening for partial conjunction hypotheses.
\newblock {\em Biometrics}, 64(4):1215--1222.

\bibitem[Birnbaum, 1954]{birnbaum1954combining}
Birnbaum, A. (1954).
\newblock Combining independent tests of significance.
\newblock {\em Journal of the American Statistical Association},
  49(267):559--574.

\bibitem[Blanchard and Roquain, 2009]{blanchard2009adaptive}
Blanchard, G. and Roquain, E. (2009).
\newblock Adaptive false discovery rate control under independence and
  dependence.
\newblock {\em J. Mach. Learn. Res.}, 10:2837--2871.

\bibitem[Fisher, 1934]{fisher1934statistical}
Fisher, R.~A. (1934).
\newblock {\em Statistical methods for research workers. Fifth Edition.}
\newblock Oliver and Boyd, Edinburgh and London.

\bibitem[Gr{\"u}nwald et~al., 2019]{grunwald2020safe}
Gr{\"u}nwald, P., de~Heide, R., and Koolen, W.~M. (2019).
\newblock Safe testing.
\newblock {\em arXiv preprint arXiv:1906.07801}.

\bibitem[Heard and Rubin-Delanchy, 2018]{heard2018choosing}
Heard, N.~A. and Rubin-Delanchy, P. (2018).
\newblock Choosing between methods of combining-values.
\newblock {\em Biometrika}, 105(1):239--246.

\bibitem[Hoang and Dickhaus, 2021a]{hoang2020usage}
Hoang, A.-T. and Dickhaus, T. (2021a).
\newblock On the usage of randomized p-values in the {S}chweder-{S}pj{\o}tvoll
  estimator.
\newblock {\em Forthcoming in the Annals of the Institute of Statistical
  Mathematics}.

\bibitem[Hoang and Dickhaus, 2021b]{hoang2021randomized}
Hoang, A.-T. and Dickhaus, T. (2021b).
\newblock Randomized p-values for multiple testing and their application in
  replicability analysis.
\newblock {\em Biometrical Journal}, early view.

\bibitem[Kocak, 2017]{kocak2017meta}
Kocak, M. (2017).
\newblock Meta-analysis of univariate p-values.
\newblock {\em Communications in Statistics-Simulation and Computation},
  46(2):1257--1265.

\bibitem[Loughin, 2004]{loughin2004systematic}
Loughin, T.~M. (2004).
\newblock A systematic comparison of methods for combining p-values from
  independent tests.
\newblock {\em Computational Statistics \& Data Analysis}, 47(3):467--485.

\bibitem[Owen et~al., 2009]{owen2009karl}
Owen, A.~B. et~al. (2009).
\newblock Karl {P}earson’s meta-analysis revisited.
\newblock {\em The Annals of Statistics}, 37(6B):3867--3892.

\bibitem[Pearson, 1938]{pearson1938probability}
Pearson, E.~S. (1938).
\newblock The probability integral transformation for testing goodness of fit
  and combining independent tests of significance.
\newblock {\em Biometrika}, 30(1/2):134--148.

\bibitem[R{\"u}schendorf, 1982]{ruschendorf1982random}
R{\"u}schendorf, L. (1982).
\newblock Random variables with maximum sums.
\newblock {\em Advances in Applied Probability}, pages 623--632.

\bibitem[Schweder and Spj{\o}tvoll, 1982]{schweder1982plots}
Schweder, T. and Spj{\o}tvoll, E. (1982).
\newblock Plots of p-values to evaluate many tests simultaneously.
\newblock {\em Biometrika}, 69(3):493--502.

\bibitem[Shaked and Shanthikumar, 2007]{shaked2007stochastic}
Shaked, M. and Shanthikumar, J.~G. (2007).
\newblock {\em Stochastic orders}.
\newblock Springer Series in Statistics. Springer, New York.

\bibitem[Stouffer et~al., 1949]{stouffer1949american}
Stouffer, S.~A., Suchman, E.~A., DeVinney, L.~C., Star, S.~A., and Williams~Jr,
  R.~M. (1949).
\newblock {\em The American soldier: Adjustment during army life.(Studies in
  social psychology in World War II), vol. 1}.
\newblock Princeton Univ. Press.

\bibitem[Vovk and Wang, 2019]{vovk2019values}
Vovk, V. and Wang, R. (2019).
\newblock E-values: Calibration, combination, and applications.
\newblock {\em Forthcoming in the Annals of Statistics}.

\bibitem[Vovk and Wang, 2020]{vovk2020combining}
Vovk, V. and Wang, R. (2020).
\newblock Combining p-values via averaging.
\newblock {\em Biometrika}, 107(4):791--808.

\bibitem[Wilkinson, 1951]{wilkinson1951statistical}
Wilkinson, B. (1951).
\newblock A statistical consideration in psychological research.
\newblock {\em Psychological bulletin}, 48(2):156.

\bibitem[Wilson, 2019]{wilson2019harmonic}
Wilson, D.~J. (2019).
\newblock The harmonic mean {$p$}-value for combining dependent tests.
\newblock {\em Proc. Natl. Acad. Sci. USA}, 116(4):1195--1200.

\end{thebibliography}

\end{document}